\documentclass[12pt,prd,showpacs,tightenlines,nofootinbib]{revtex4}
\usepackage{bm}
\usepackage{graphics}
\usepackage{rotating}
\usepackage{epsfig}
\begin{document}

\title{Excited heavy tetraquarks with hidden charm}
\author{D. Ebert$^{1,2}$, R. N. Faustov$^{3}$  and V. O. Galkin$^{3}$}
\affiliation{$^1$ Bogoliubov Laboratory for Theoretical Physics, Joint Institute for Nuclear Research, Dubna,
  R-141980, Russia\\
$^2$ Institut f\"ur Physik, Humboldt--Universit\"at zu Berlin,
Newtonstr. 15, D-12489  Berlin, Germany\\
$^3$ Dorodnicyn Computing Centre, Russian Academy of Sciences,
  Vavilov Str. 40, 119991 Moscow, Russia}

\begin{abstract}
 The masses of the excited heavy tetraquarks with hidden
charm  are calculated within  the relativistic
diquark-antidiquark picture. The dynamics of the light quark in a
heavy-light diquark is treated completely relativistically. The
diquark structure is taken into account by calculating
the diquark-gluon form factor. New experimental data on charmonium-like states
above open charm threshold are discussed. The obtained results
indicate that  $X(3872)$, $Y(4260)$, $Y(4360)$, $Z(4248)$, $Z(4433)$ and
$Y(4660)$ could be tetraquark states with hidden charm.  
\end{abstract}

\pacs{12.40.Yx, 14.40.Gx, 12.39.Ki}

\maketitle

Recently, significant experimental progress has been achieved in
charmonium spectroscopy. Several  new 
states, such as $X(3872)$, $Y(4260)$, $Y(4360)$, $Y(4660)$, $Z(4248)$,
$Z(4430)$, etc., were 
observed \cite{pakhlova} which cannot be simply accommodated in the 
quark-antiquark ($c\bar c$) picture. These  states and especially the
charged ones
can be considered as indications of the possible existence of
exotic multiquark states \cite{jm,bk}. In  our  papers
\cite{tetr1,tetr2} we calculated masses of the ground state 
heavy tetraquarks in  the framework of the relativistic quark
model based on the quasipotential approach in quantum chromodynamics.
Here we extend this analysis to the consideration of the
excited tetraquark states with hidden charm.
As previously, we use the  diquark-antidiquark
picture to reduce a complicated relativistic 
four-body problem to the subsequent two more simple two-body
problems. The first step consists in the calculation of the masses, wave
functions and form factors of the diquarks, composed from light and heavy
quarks. At the second step, a heavy tetraquark is considered to be a
bound diquark-antidiquark system. It is 
important to emphasize that we do not consider the diquark as a point
particle but explicitly take into account its structure by calculating
the form factor of the diquark-gluon interaction in terms of the
diquark wave functions.

In the quasipotential approach and diquark-antidiquark picture of
heavy tetraquarks the interaction of two quarks in a diquark and
the diquark-antidiquark interaction in a tetraquark are described
by the diquark wave function ($\Psi_{d}$) of the bound quark-quark
state and by the tetraquark wave function ($\Psi_{T}$) of the
bound diquark-antidiquark state, respectively. These wave functions satisfy the
quasipotential equation of the Schr\"odinger type \cite{efg}
\begin{equation}
\label{quas}
{\left(\frac{b^2(M)}{2\mu_{R}}-\frac{{\bf
p}^2}{2\mu_{R}}\right)\Psi_{d,T}({\bf p})} =\int\frac{d^3 q}{(2\pi)^3}
 V({\bf p,q};M)\Psi_{d,T}({\bf q}),
\end{equation}
where the relativistic reduced mass is
\begin{equation}
\mu_{R}=\frac{E_1E_2}{E_1+E_2}=\frac{M^4-(m^2_1-m^2_2)^2}{4M^3},
\end{equation}
and $E_1$, $E_2$ are given by
\begin{equation}
\label{ee}
E_1=\frac{M^2-m_2^2+m_1^2}{2M}, \quad E_2=\frac{M^2-m_1^2+m_2^2}{2M}.
\end{equation}
Here, $M=E_1+E_2$ is the bound-state mass (diquark or tetraquark),
$m_{1,2}$ are the masses of quarks ($q$ and $Q$) which form
the diquark or of the diquark ($d$) and antiquark ($\bar d'$) which
form the heavy tetraquark ($T$), and ${\bf p}$ is their relative
momentum. In the center-of-mass system the relative momentum
squared on mass shell reads
\begin{equation}
{b^2(M) }
=\frac{[M^2-(m_1+m_2)^2][M^2-(m_1-m_2)^2]}{4M^2}.
\end{equation}

The kernel $V({\bf p,q};M)$ in Eq.~(\ref{quas}) is the
quasipotential operator of the quark-quark or diquark-antidiquark
interaction. It is constructed with the help of the off-mass-shell
scattering amplitude, projected onto the positive-energy states.
In the following analysis we closely follow the similar
construction of the quark-antiquark interaction in mesons which
were extensively studied in our relativistic quark model
\cite{efg,egf}. For the quark-quark interaction in a diquark we
use the relation $V_{qq}=V_{q\bar q}/2$ arising under the
assumption of an octet structure of the interaction from the
difference in the $qq$ and $q\bar q$ colour
states. An important role in this construction is played by the
Lorentz structure of the confining interaction. In our analysis of
mesons, while constructing the quasipotential of the
quark-antiquark interaction, we assumed that the effective
interaction is the sum of the usual one-gluon exchange term and a
mixture of long-range vector and scalar linear confining
potentials, where the vector confining potential contains the
Pauli term. We use the same conventions for the construction of
the quark-quark and diquark-antidiquark interactions in the
tetraquark. The quasipotential is then defined as follows
\cite{efgm,egf}.

(a) For the quark-quark  ($Qq$) interactions,
$V({\bf p,q};M)$ reads
 \begin{equation}
\label{qpot}
V({\bf p,q};M)=\bar{u}_{1}(p)\bar{u}_{2}(-p){\cal V}({\bf p}, {\bf
q};M)u_{1}(q)u_{2}(-q),
\end{equation}
with
\[
{\cal V}({\bf p,q};M)=\frac12\left[\frac43\alpha_sD_{ \mu\nu}({\bf
k})\gamma_1^{\mu}\gamma_2^{\nu}+ V^V_{\rm conf}({\bf k})
\Gamma_1^{\mu}({\bf k})\Gamma_{2;\mu}(-{\bf k})+
 V^S_{\rm conf}({\bf k})\right].
\]
Here, $\alpha_s$ is the QCD coupling constant; $D_{\mu\nu}$ is the
gluon propagator in the Coulomb gauge,
\begin{equation}
D^{00}({\bf k})=-\frac{4\pi}{{\bf k}^2}, \quad D^{ij}({\bf k})=
-\frac{4\pi}{k^2}\left(\delta^{ij}-\frac{k^ik^j}{{\bf k}^2}\right),
\quad D^{0i}=D^{i0}=0,
\end{equation}
and ${\bf k=p-q}$; $\gamma_{\mu}$ and $u(p)$ are the Dirac
matrices and spinors,
\begin{equation}
\label{spinor}
u^\lambda({p})=\sqrt{\frac{\epsilon(p)+m}{2\epsilon(p)}}
\left(\begin{array}{c} 1\\
\displaystyle\frac{\mathstrut\bm{\sigma}\cdot{\bf p}}
{\mathstrut\epsilon(p)+m}
\end{array}\right)
\chi^\lambda,
\end{equation}
with $\epsilon(p)=\sqrt{{\bf p}^2+m^2}$.

The effective long-range vector vertex of the quark is
defined \cite{egf} by
\begin{equation}
\Gamma_{\mu}({\bf k})=\gamma_{\mu}+
\frac{i\kappa}{2m}\sigma_{\mu\nu}\tilde k^{\nu}, \qquad \tilde
k=(0,{\bf k}),
\end{equation}
where $\kappa$ is the Pauli interaction constant characterizing
the anomalous chromomagnetic moment of quarks. In configuration
space the vector and scalar confining potentials in the
nonrelativistic limit \cite{efgr} reduce to
\begin{eqnarray}
V^V_{\rm conf}(r)&=&(1-\varepsilon)V_{\rm
conf}(r),\nonumber\\[1ex] V^S_{\rm conf}(r)& =&\varepsilon V_{\rm
conf}(r),
\end{eqnarray}
with
\begin{equation}
V_{\rm conf}(r)=V^S_{\rm conf}(r)+
V^V_{\rm conf}(r)=Ar+B,
\end{equation}
where $\varepsilon$ is the mixing coefficient.

(b) For the diquark-antidiquark ($d\bar d'$) interaction, $V({\bf
p,q};M)$ is given by
\begin{eqnarray}
\label{dpot} V({\bf p,q};M)&=&\frac{\langle
d(P)|J_{\mu}|d(Q)\rangle} {2\sqrt{E_dE_d}} \frac43\alpha_sD^{
\mu\nu}({\bf k})\frac{\langle d'(P')|J_{\nu}|d'(Q')\rangle}
{2\sqrt{E_{d'}E_{d'}}}\nonumber\\[1ex]
&&+\psi^*_d(P)\psi^*_{d'}(P')\left[J_{d;\mu}J_{d'}^{\mu} V_{\rm
conf}^V({\bf k})+V^S_{\rm conf}({\bf
k})\right]\psi_d(Q)\psi_{d'}(Q'),
\end{eqnarray}
where $\langle
d(P)|J_{\mu}|d(Q)\rangle$ is the vertex of the
diquark-gluon interaction which takes into account the finite size of
the diquark and is discussed
below
$\Big[$$P^{(')}=(E_{d^{(')}},\pm{\bf p})$ and
$Q^{(')}=(E_{d^{(')}},\pm{\bf q})$,
$E_d=(M^2-M_{d'}^2+M_d^2)/(2M)$ and $E_{d'}=(M^2-M_d^2+M_{d'}^2)/(2M)$
$\Big]$.

The diquark state in the confining part of the diquark-antidiquark
quasipotential (\ref{dpot}) is described by the wave functions
\begin{equation}
 \label{eq:ps}
 \psi_d(p)=\left\{\begin{array}{ll}1 &\qquad \text{for a scalar
 diquark,}\\[1ex]
\varepsilon_d(p) &\qquad \text{for an axial-vector diquark,}
\end{array}\right.
\end{equation}
where the four-vector
\begin{equation}\label{pv}
\varepsilon_d(p)=\left(\frac{(\bm{\varepsilon}_d\cdot{\bf
p})}{M_d},\bm{\varepsilon}_d+ \frac{(\bm{\varepsilon}_d\cdot{\bf
p}){\bf
 p}}{M_d(E_d(p)+M_d)}\right), \qquad \varepsilon^\mu_d(p) p_\mu=0,
\end{equation}
is the polarization vector of the axial-vector diquark with
momentum ${\bf p}$, $E_d(p)=\sqrt{{\bf p}^2+M_d^2}$, and
$\varepsilon_d(0)=(0,\bm{\varepsilon}_d)$ is the polarization
vector in the diquark rest frame. The effective long-range vector
vertex of the diquark can be presented in the form
\begin{equation}
 \label{eq:jc}
 J_{d;\mu}=\left\{\begin{array}{ll}
 \frac{\displaystyle (P+Q)_\mu}{\displaystyle
 2\sqrt{E_dE_d}}&\qquad \text{ for a scalar diquark,}\\[3ex]
-\; \frac{\displaystyle (P+Q)_\mu}{\displaystyle2\sqrt{E_dE_d}}
 +\frac{\displaystyle i\mu_d}{\displaystyle 2M_d}\Sigma_\mu^\nu
\tilde k_\nu
 &\qquad \text{ for an axial-vector diquark.}\end{array}\right.
\end{equation}
Here, the antisymmetric tensor
$\Sigma_\mu^\nu$ is defined by
\begin{equation}
 \label{eq:Sig}
 \left(\Sigma_{\rho\sigma}\right)_\mu^\nu=-i(g_{\mu\rho}\delta^\nu_\sigma
 -g_{\mu\sigma}\delta^\nu_\rho),
\end{equation}
and the axial-vector diquark spin ${\bf S}_d$ is given by
$(S_{d;k})_{il}=-i\varepsilon_{kil}$; $\mu_d$ is the total
chromomagnetic moment of the axial-vector diquark.

The constituent quark masses $m_c=1.55$ GeV,
$m_u=m_d=0.33$ GeV, $m_s=0.5$ GeV and the parameters of the linear
potential $A=0.18$ GeV$^2$ and $B=-0.3$~GeV have values
typical in quark models. The value of the mixing coefficient of
vector and scalar confining potentials $\varepsilon=-1$ has been
determined from the consideration of charmonium radiative decays
\cite{efg} and the heavy-quark expansion \cite{fg}. The universal
Pauli interaction constant $\kappa=-1$ has been fixed from the
analysis of the fine splitting of heavy quarkonia ${ }^3P_J$ -
states \cite{efg}. In this case, the long-range chromomagnetic
interaction of quarks vanishes in accordance with the flux-tube
model.

At the first step, we calculate the masses and form factors of the
 heavy-light
diquark. As it is well known, the light quarks are highly
relativistic, which makes the $v/c$ expansion inapplicable and thus,
a completely relativistic treatment of the light quark dynamics is
required. To achieve this goal,  we closely follow our consideration
of diquarks in heavy baryons
and adopt the same procedure to make the relativistic
potential local by replacing
$\epsilon_{1,2}(p)=\sqrt{m_{1,2}^2+{\bf p}^2}\to E_{1,2}=(M^2-m_{2,1}^2+m_{1,2}^2)/2M$. 
Solving numerically the quasipotential equation (\ref{quas}) with the
complete relativistic potential,  which depends on the
diquark mass in a complicated highly nonlinear way \cite{hbar}, we get
the diquark masses and wave functions. In order to determine the
diquark interaction with the gluon field, which 
takes into account the diquark structure, we
calculate the corresponding matrix element of the quark
current between diquark states. Such calculation leads to the
emergence of the form factor $F(r)$ entering the vertex of the
diquark-gluon interaction \cite{hbar}. This form factor is expressed
through the overlap integral of the diquark wave functions. Our estimates show that this form factor can be approximated  with a
high accuracy by the expression 
\begin{equation}
  \label{eq:fr}
  F(r)=1-e^{-\xi r -\zeta r^2}.
\end{equation}
The values of the masses and parameters $\xi$ and $\zeta$ for heavy-light
scalar diquark $[\cdots]$ and axial vector diquark $\{\cdots\}$ ground states are
given in Table~\ref{tab:dmass}.

\begin{table}
  \caption{Masses $M$ and form factor  parameters of charmed
    diquarks. $S$ and $A$ 
    denote scalar and axial vector diquarks which are antisymmetric $[\cdots]$ and
    symmetric $\{\cdots\}$ in flavour, respectively. }
  \label{tab:dmass}
\begin{ruledtabular}
\begin{tabular}{ccccc}
Quark& Diquark&   $M$ &$\xi$ & $\zeta$
 \\
content &type & (MeV)& (GeV)& (GeV$^2$)  \\
\hline
$[c,q]$& $S$ & 1973& 2.55 &0.63  \\
$\{c,q\}$& $A$ & 2036& 2.51 &0.45  \\
$[c,s]$ & $S$& 2091& 2.15 & 1.05  \\
$\{c,s\}$& $A$ & 2158&2.12& 0.99 
  \end{tabular}
\end{ruledtabular}
\end{table}

At the second step, we calculate the masses of heavy tetraquarks 
considered as the bound states of a heavy-light diquark and
antidiquark. For the
potential of the diquark-antidiquark interaction (\ref{dpot}) we
get \cite{tetr2}  
\begin{eqnarray}
 \label{eq:pot}
 V(r)&=& \hat V_{\rm Coul}(r)+V_{\rm conf}(r)+\frac12\Biggl\{\left[
   \frac1{E_1(E_1+M_1)}+\frac1{E_2(E_2+M_2)}\right]
\frac{\hat V'_{\rm Coul}(r)}r 
-\Biggl[\frac1{M_1(E_1+M_1)}\nonumber\\[1ex]
&& +\frac1{M_2(E_2+M_2)}\Biggr]
\frac{V'_{\rm conf}(r)}r +\frac{\mu_d}2
\left(\frac1{M_1^2}+\frac1{M_2^2}\right)\frac{V'^V_{\rm conf}(r)}r\Biggr\}
{\bf L}\cdot ({\bf
S}_1+{\bf S}_2 )\nonumber\\[1ex]
&&+\frac12\Biggl\{\left[
   \frac1{E_1(E_1+M_1)}-\frac1{E_2(E_2+M_2)}\right]
\frac{\hat V'_{\rm Coul}(r)}r 
-\left[\frac1{M_1(E_1+M_1)}-\frac1{M_2(E_2+M_2)}\right]\nonumber\\[1ex]
&& \times
\frac{V'_{\rm conf}(r)}r +\frac{\mu_d}2
\left(\frac1{M_1^2}-\frac1{M_2^2}\right)\frac{V'^V_{\rm conf}(r)}r\Biggl\}
{\bf L}\cdot ({\bf
S}_1-{\bf S}_2 )\nonumber\\[1ex]
&&+\frac1{E_1E_2}\Biggl\{{\bf
 p}\left[\hat V_{\rm Coul}(r)+V^V_{\rm conf}(r)\right]{\bf p} -\frac14
\Delta V^V_{\rm conf}(r)+ \hat V'_{\rm Coul}(r)\frac{{\bf
 L}^2}{2r}\nonumber\\[1ex]
&& +\frac1{r}\left[\hat V'_{\rm
Coul}(r)+\frac{\mu_d}4\left(\frac{E_1}{M_1}
+\frac{E_2}{M_2}\right)V'^V_{\rm conf}(r)\right]{\bf L}\cdot ({\bf
S}_1+{\bf S}_2)\nonumber\\[1ex] &&
+\frac{\mu_d}4\left(\frac{E_1}{M_1}
-\frac{E_2}{M_2}\right)\frac{V'^V_{\rm conf}(r)}{r}{\bf
L}\cdot({\bf S}_1-{\bf S}_2)\nonumber\\[1ex] &&
+\frac13\left[\frac1{r}{\hat V'_{\rm Coul}(r)}-\hat V''_{\rm
Coul}(r) +\frac{\mu_d^2}4\frac{E_1E_2}{M_1M_2}
\left(\frac1{r}{V'^V_{\rm conf}(r)}-V''^V_{\rm
 conf}(r)\right)\right]\nonumber\\[1ex]
&&\times
 \left[\frac3{r^2}({\bf S}_1\cdot{\bf r}) ({\bf
 S}_2\cdot{\bf r})-
{\bf S}_1\cdot{\bf S}_2\right]\nonumber\\[1ex] &&
+\frac23\left[\Delta \hat V_{\rm
Coul}(r)+\frac{\mu_d^2}4\frac{E_1E_2}{M_1M_2} \Delta V^V_{\rm
conf}(r)\right]{\bf S}_1\cdot{\bf S}_2\Biggr\},
\end{eqnarray}
where $$\hat V_{\rm Coul}(r)=-\frac{4}{3}\alpha_s
\frac{F_1(r)F_2(r)}{r}$$ is the Coulomb-like one-gluon exchange
potential which takes into account the finite sizes of the diquark
and antidiquark through corresponding form factors $F_{1,2}(r)$.
Here, ${\bf S}_{1,2}$ and ${\bf L}$ are the spin operators of
diquark and antidiquark and the operator of the relative orbital
angular momentum.  In the
following we choose the total chromomagnetic moment of the
axial-vector diquark $\mu_d=0$. Such a choice appears to be
natural, since the long-range chromomagnetic interaction of
diquarks proportional to $\mu_d$ then also vanishes in accordance
with the flux-tube model.

In the diquark-antidiquark picture of heavy tetraquarks
both scalar $S$ (antisymmetric in flavour
$(Qq)_{S=0}=[Qq]$) and axial vector $A$ (symmetric in flavour
$(Qq)_{S=1}=\{Qq\}$) diquarks are considered. Therefore we get the
following structure of the $(Qq)(\bar Q\bar q')$  ground ($1S$) states
($C$ is defined only for $q=q'$): 
\begin{itemize}
\item Two states with $J^{PC}=0^{++}$:
\begin{eqnarray*}
&&X(0^{++})=(Qq)_{S=0}(\bar Q\bar q')_{S=0}\\
&&X(0^{++}{}')=(Qq)_{S=1}(\bar Q\bar q')_{S=1}
\end{eqnarray*}
\item Three states with $J=1$:
\begin{eqnarray*}
&&X(1^{++})=\frac1{\sqrt{2}}[(Qq)_{S=1}(\bar Q\bar q')_{S=0}+(Qq)_{S=0}(\bar Q\bar
  q')_{S=1}]\\
&&X(1^{+-})=\frac1{\sqrt{2}}[(Qq)_{S=0}(\bar Q\bar q')_{S=1}-(Qq)_{S=1}(\bar Q\bar
  q')_{S=0}]\\
&&X(1^{+-}{}')=(Qq)_{S=1}(\bar Q\bar q')_{S=1}
\end{eqnarray*}
\item  One state with $J^{PC}=2^{++}$:
$$X(2^{++})=(Qq)_{S=1}(\bar Q\bar q')_{S=1}.$$
\end{itemize}
The orbitally excited ($1P,1D\dots$) states are
constructed analogously. As we find, a very rich spectrum of tetraquarks
emerges. However the number of states in the considered
diquark-antidiquark picture is significantly less than in the genuine
four-quark approach.

The diquark-antidiquark model of heavy tetraquarks predicts
 the existence of a flavour 
$SU(3)$ nonet of states with hidden
charm or beauty ($Q=c,b$): four tetraquarks
[$(Qq)(\bar Q\bar q)$, $q=u,d$] with neither open 
or hidden strangeness, which have
electric charges 0 or $\pm 1$ and isospin 0 or 1; 
four tetraquarks [$(Qs)(\bar Q\bar q)$
and  $(Qq)(\bar Q\bar s)$, $q=u,d$] with open strangeness ($S=\pm 1$),
which have electric charges 0 or $\pm 1$ and isospin $\frac12$; 
one tetraquark
$(Qs)(\bar Q\bar s)$ with hidden strangeness and zero electric
charge. 
Since we neglect  in our model the mass difference of $u$ and
$d$ quarks and electromagnetic interactions, the corresponding tetraquarks
will be degenerate in mass. A more 
detailed analysis \cite{mppr}
predicts that the tetraquark mass differences can be of a few MeV so
that the
isospin invariance is broken for the $(Qq)(\bar Q\bar q)$ mass
eigenstates and thus in their strong decays.  
The (non)observation of such states will be a crucial test of the
tetraquark model.

\begin{table}
  \caption{Masses of charm diquark-antidiquark ground ($1S$) states
    (in MeV) calculated in \cite{tetr1}. $S$ and $A$
    denote scalar and axial vector diquarks. }
  \label{tab:cmass}
\begin{ruledtabular}
\begin{tabular}{ccccc}
State& Diquark &
\multicolumn{3}{l}{\underline{\hspace{3.4cm}Mass\hspace{3.4cm}}} 
\hspace{-5.5cm} \\
$J^{PC}$ & content& $cq\bar c\bar q$ &$cs\bar c\bar s$ & $ cq\bar c\bar s$ \\
\hline
$0^{++}$ & $S\bar S$ & 3812 & 4051 & 3922\\
$1^{+\pm}$ & $(S\bar A\pm \bar S A)/\sqrt2$& 3871& 4113 & 3982\\
$0^{++}$& $A\bar A$ & 3852 & 4110& 3967\\
$1^{+-}$& $A\bar A$ & 3890 & 4143& 4004\\
$2^{++}$& $A\bar A$ & 3968 & 4209&4080\\
 \end{tabular}
\end{ruledtabular}
\end{table}

\begin{table}
  \caption{Thresholds for open charm decays and nearby hidden-charm
    thresholds.} 
  \label{tab:cthr}
\begin{ruledtabular}
\begin{tabular}{cccccc}
Channel& Threshold (MeV)&Channel& Threshold (MeV)&Channel& Threshold
(MeV)\\ 
\hline
$D^0\bar D^0$& 3729.4 &$D_s^+ D_s^-$& 3936.2&$D^0 D_s^\pm$& 3832.9\\
$D^+D^-$& 3738.8& $\eta' J/\psi$& 4054.7& $D^\pm D_s^\mp$ & 3837.7\\
$D^0\bar D^{*0}$ & 3871.3& $D_s^\pm D_s^{*\mp}$& 4080.0&$D^{*0} D_s^\pm$ &
3975.0\\
$\rho J/\psi$& 3872.7& $\phi J/\psi$ & 4116.4&$D^{0}D^{*\pm}_s$ &
3976.7\\
$D^\pm D^{*\mp}$ &3879.5 &$D^{*+}_sD^{*-}_s$& 4223.8& $K^{*\pm}J/\psi$ &
3988.6\\
$\omega J/\psi$& 3879.6 & & &  $K^{*0}J/\psi$ & 3993.0\\
$D^{*0}\bar D^{*0}$ & 4013.6 & & &$D^{*0} D_s^{*\pm}$ & 4118.8   
\end{tabular}
\end{ruledtabular}
\end{table}

The calculated  masses of the heavy tetraquark ground ($1S$) states 
and the corresponding open charm  thresholds are shown in
Tables~~\ref{tab:cmass}, \ref{tab:cthr}. Note that most of the
tetraquark states were predicted to lie 
either above or only slightly below corresponding open charm
thresholds. In Table~\ref{tab:ecmass},~\ref{tab:ecmass1} we give our predictions for the
orbitally and radially excited tetraquark states with hidden
charm. Excitations only of the diquark-antidiquark system are
considered. A very rich spectrum of excited tetraquark
states is obtained.

\begin{table}
  \caption{Masses of charm diquark-antidiquark excited  $1P$, $2S$ states
     (in MeV).  $S$ and $A$
    denote scalar and axial vector diquarks; $\cal S$ is the total
    spin of the diquark and antidiquark. ($C$ is defined only for $q=q'$).}
  \label{tab:ecmass}
\begin{ruledtabular}
\begin{tabular}{cccccc}
State& Diquark & &
\multicolumn{3}{l}{\underline{\hspace{2.8cm}Mass\hspace{2.8cm}}} 
\hspace{-5.5cm} \\
$J^{PC}$ & content&$\cal S$& $cq\bar c\bar q$ &$cs\bar c\bar s$ & $cq\bar c\bar s$ \\
\hline
$1P$\\
$1^{--}$ & $S\bar S$& 0 & 4244& 4466& 4350\\
$0^{-\pm}$ & $(S\bar A\pm \bar S A)/\sqrt2$&1&4269 & 4499&4381 \\
$1^{-\pm}$ & $(S\bar A\pm \bar S A)/\sqrt2$&1&4284 &4514& 4396 \\
$2^{-\pm}$ & $(S\bar A\pm \bar S A)/\sqrt2$&1&4315 &4543 &4426 \\
$1^{--}$& $A\bar A$ &0&4350& 4582& 4461\\
$0^{-+}$& $A\bar A$ & 1&4304 &4540& 4419\\
$1^{-+}$& $A\bar A$ & 1& 4345 &4578& 4458\\
$2^{-+}$& $A\bar A$ & 1& 4367 & 4598& 4478\\
$1^{--}$& $A\bar A$ & 2&4277 &4515& 4393\\
$2^{--}$& $A\bar A$ & 2& 4379& 4610& 4490\\
$3^{--}$& $A\bar A$ & 2& 4381& 4612& 4492\\
 $2S$\\
$0^{++}$ & $S\bar S$ & 0 & 4375 &4604&4481\\
$1^{+\pm}$ & $(S\bar A\pm \bar S A)/\sqrt2$& 1& 4431 & 4665& 4542\\
$0^{++}$& $A\bar A$ & 0 & 4434 &4680& 4547\\
$1^{+-}$& $A\bar A$ & 1 & 4461 &4703&4572\\
$2^{++}$& $A\bar A$ & 2 & 4515& 4748& 4625\\
 \end{tabular}
\end{ruledtabular}
\end{table}

\begin{table}
  \caption{Masses of charm diquark-antidiquark excited  $1D$, $2P$ states
    (in MeV). $S$ and $A$
    denote scalar and axial vector diquarks; $\cal S$ is the total
    spin of the diquark and antidiquark. }
  \label{tab:ecmass1}
\begin{ruledtabular}
\begin{tabular}{cccccc}
State& Diquark & &
\multicolumn{3}{l}{\underline{\hspace{2.8cm}Mass\hspace{2.8cm}}} 
\hspace{-5.5cm} \\
$J^{PC}$ & content&$\cal S$& $cq\bar c\bar q$ &$cs\bar c\bar s$ & $ cq\bar c\bar s$ \\
\hline
$1D$\\
$2^{++}$ & $S\bar S$& 0 & 4506& 4728& 4611\\
$1^{+\pm}$ & $(S\bar A\pm \bar S A)/\sqrt2$&1&4553 &4779& 4663\\
$2^{+\pm}$ & $(S\bar A\pm \bar S A)/\sqrt2$&1&4559 &4785& 4670\\
$3^{+\pm}$ & $(S\bar A\pm \bar S A)/\sqrt2$&1&4570 &4794& 4680\\
$2^{++}$& $A\bar A$ &0&4617& 4847&4727\\
$1^{+-}$& $A\bar A$ & 1&4604 &4835&4714\\
$2^{+-}$& $A\bar A$ & 1& 4616&4846&4726\\
$3^{+-}$& $A\bar A$ & 1& 4624&4852& 4733\\
$0^{++}$& $A\bar A$ & 2&4582 &4814& 4692\\
$1^{++}$& $A\bar A$ & 2&4593& 4825& 4703\\
$2^{++}$& $A\bar A$ & 2& 4610 & 4841& 4720\\
$3^{++}$& $A\bar A$ & 2& 4627 &4855& 4736\\
$4^{++}$& $A\bar A$ & 2&4628& 4856& 4738\\
$2P$\\
$1^{--}$ & $S\bar S$& 0 & 4666& 4884& 4767\\
$0^{-\pm}$ & $(S\bar A\pm \bar S A)/\sqrt2$&1&4684 & 4909&4792\\
$1^{-\pm}$ & $(S\bar A\pm \bar S A)/\sqrt2$&1&4702 & 4926& 4810\\
$2^{-\pm}$ & $(S\bar A\pm \bar S A)/\sqrt2$&1&4738 & 4960& 4845\\
$1^{--}$& $A\bar A$ &0&4765& 4991&4872\\
$0^{-+}$& $A\bar A$ & 1&4715& 4946& 4826\\
$1^{-+}$& $A\bar A$ & 1& 4760 &4987& 4867\\
$2^{-+}$& $A\bar A$ & 1& 4786& 5011& 4892\\
$1^{--}$& $A\bar A$ & 2&4687&4920&4799\\
$2^{--}$& $A\bar A$ & 2& 4797 &5022&4903\\
$3^{--}$& $A\bar A$ & 2& 4804 &5030&4910\\
 \end{tabular}
\end{ruledtabular}
\end{table}

\begin{table}
  \caption{Comparison of theoretical predictions for  the masses of
    the ground and excited
   charm diquark-antidiquark states   (in MeV) and
   possible experimental candidates.}
  \label{tab:cemass}
\begin{ruledtabular}
\begin{tabular}{ccccccc}
State&Diquark&
\multicolumn{3}{l}{\underline{\hspace{2.5cm}Theory\hspace{2.5cm}}}
& \multicolumn{2}{l}{\underline{\hspace{1.9cm}Experiment
    \hspace{1.9cm}}} 
\hspace{-1.5cm}  \\
$J^{PC}$&content &EFG & \cite{mppr,mpr,mprZ} &\cite{mpprY} ($cs\bar c\bar s$) 
&state& mass\\
\hline
$1S$\\
$0^{++}$&$S\bar S$ & 3812 & 3723& & &\\
$1^{++}$&$(S\bar A+ \bar S A)/\sqrt2$ & 3871& 3872$^\dag$&
&$\left\{\begin{array}{l} X{(3872)}\\ X{(3876)}\end{array}\right.$ 
&$\left\{\begin{array}{l}{3871.4}\pm{0.6} \ [1] \\
{3875.2}\pm{0.7}^{+0.9}_{-1.8} \  [1]\end{array}\right.$ \\
$1^{+-}$&$(S\bar A- \bar S A)/\sqrt2$ & 3871& 3754&& &\\
$0^{++}$&$A\bar A$& 3852 & 3832&& &\\
$1^{+-}$&$A\bar A$& 3890 & 3882&& &\\
$2^{++}$&$A\bar A$& 3968 &
3952&&$Y$(3943)&$\left\{\begin{array}{l}3943\pm11\pm13 \ [16] 
\\3914.3^{+4.1}_{-3.8} \ [17]\end{array}\right.$\\
$1P$\\
$1^{--}$&$S\bar S$&4244 & &4330$\pm$70&$Y$(4260)
&$\left\{\begin{array}{l}{4259}\pm{8}^{+2}_{-6} \ [18] \\
   {4247}\pm{12}^{+17}_{-32}\ [19]\end{array}\right.$\\
{$\begin{array}{l}1^{-}\\
  0^{-} \end{array}$}&{$\begin{array}{c}S\bar S\\
  (S\bar A\pm \bar S A)/\sqrt2 \end{array}$}&$\left.\begin{array}{l}4244
  \\4267\end{array}\right\}$&&&$Z(4248)$&$4248^{+44+180}_{-29-35}$
[20] \\
\cr
$\begin{array}{l}1^{--}\\1^{--} \end{array}$&$\begin{array}{c}(S\bar
  A- \bar S A)/\sqrt2\\ A\bar
  A \end{array}$&$\left.\begin{array}{l}4284\\4277 \end{array}\right\}$& &&$Y$(4260) &
4284$^{+17}_{-16}\pm$4 [21]\\
$1^{--}$&$A\bar A$& 4350& &  &$Y$(4360) &
$\left\{\begin{array}{l}4361\pm9\pm9 \ [22]
\\4324\pm24 \ [23] \end{array}\right.$\\
$2S$\\
$\begin{array}{l}1^{+}\\0^{+}\end{array}$&$\begin{array}{c}(S\bar A\pm \bar S A)/\sqrt2\\A\bar A \end{array}$&$\left.\begin{array}{l}4431\\4434\end{array}\right\}$& &
&$Z$(4430)&4433$\pm$4$\pm$2 [24]\\
$1^{+}$&$A\bar A$&4461& $\sim$ 4470\\
$2P$\\
$1^{--}$&$S\bar S$&4666& & & $\left\{\begin{array}{l}Y(4660)\\X(4630)\end{array}\right.$&
$\left\{\begin{array}{l}4664\pm11\pm5 \ [22]
 \\ 4634^{+8+5}_{-7-8} \ [25] \end{array}\right.$\\ 
\end{tabular}
 \end{ruledtabular}
\flushleft{${}^\dag$ input}
\end{table}

In Table~\ref{tab:cemass} we compare our results (EFG) for
the masses of the ground and excited charm 
diquark-antidiquark bound states with the predictions of
Refs.~\cite{mppr,mpr,mprZ,mpprY} and with the masses of the recently observed
highly-excited charmonium-like states \cite{pakhlova,3943belle,3943babar,4260babar,4260belle,bellez1z2,4260cleo,4360belle,4360babar,4430belle,belley}. 
We  assume that the excitations  occur only between the bound 
diquark and antidiquark. Possible excitations of diquarks are
not considered. Our calculation of the heavy baryon masses supports
such a scheme \cite{hbar}.
In this table we give our predictions only for some of the masses of the
orbitally and radially excited states for which possible experimental
candidates are observed. 
The differences in some of the presented theoretical mass values can
be attributed to the substantial distinctions in the used
approaches. We describe the diquarks dynamically as 
quark-quark bound 
systems and  calculate their masses and form factors, while in
Refs.\cite{mppr,mpr,mprZ,mpprY}  they
are treated only phenomenologically. Then we consider the tetraquark
as purely the 
diquark-antidiquark bound system.  In distinction, Maini et al. 
consider a
hyperfine interaction between all quarks  which, e.g., causes the
splitting of $1^{++}$ and $1^{+-}$ states arising from the $SA$
diquark-antidiquark compositions.  
From Table~\ref{tab:cemass} we see that our dynamical 
calculation supports  the assumption \cite{mppr} that $X(3872)$ can be
the axial vector 
$1^{++}$ tetraquark state composed from the scalar and axial vector
diquark and antidiquark in the relative $1S$ state. Recent Belle and
BaBar results indicate the existence of a second $X(3875)$ particle a
few MeV above $X(3872)$. This state could be naturally identified
with the second neutral particle predicted by the tetraquark model \cite{mpr}.  
On the other hand, in our model 
the lightest scalar $0^{++}$ tetraquark is
predicted to be above the open charm threshold $D\bar D$
and thus to be broad, while in the model \cite{mppr} it lies a
few MeV below this threshold, and thus is predicted to be narrow. Our
$2^{++}$ tetraquark also lies higher than the one in Ref.\cite{mppr},
thus making the interpretation of this state as $Y(3943)$ less
probable, especially if one averages the original Belle result with the recent
BaBar value which is somewhat lower.  

The recent discovery in the initial state radiation at $B$-factories of the $Y(4260)$, $Y(4360)$ and $Y(4660)$
indicates an overpopulation of the expected charmonium $1^{--}$
states
\cite{pakhlova,4260babar,4260belle,4260cleo,4360belle,4360babar}. 
Maini et al. \cite{mpprY} argue that
$Y(4260)$ is the  $1^{--}$ $1P$ state of the charm-strange
diquark-antidiquark tetraquark.
We find that $Y(4260)$ cannot be interpreted in this way, since the mass
of such $([cs]_{S=0}[\bar c\bar s]_{S=0})$ tetraquark  is found to
be $\sim 200$ MeV higher. A more natural
tetraquark interpretation could be the $1^{--}$ $1P$ state  $([cq]_{S=0}[\bar
c\bar q]_{S=0})$ ($S\bar S$) which mass is predicted in our model to be 
close to the mass of  $Y(4260)$ (see Table~\ref{tab:cemass}). 
Then the $Y(4260)$ would decay dominantly into $D\bar D$ pairs.
The other possible interpretations of  $Y(4260)$ are the $1^{--}$ $1P$ states of
$(S\bar A- \bar S A)/\sqrt2$ and $A\bar A$ tetraquarks which predicted masses
have close values. These additional tetraquark states could be
responsible for the mass difference of $Y(4260)$  observed in different
decay channels. As we see from Table~\ref{tab:cemass}, the recently
discovered resonances  $Y(4360)$ and 
$Y(4660)$ in the $e^+e^-\to\pi^+\pi^-\psi'$ cross section can be
interpreted as the excited $1^{--}$ $1P$ ($A\bar A$) and $2P$ ($S\bar S$)
tetraquark states, respectively.  The peak $X(4630)$ very
recently observed by Belle in $e^+e^-\to\Lambda^+_c\Lambda^-_c$
\cite{belley} is
consistent with a $1^{--}$ resonance  $Y(4660)$ and therefore has the same
interpretation in our model.

Recently the Belle Collaboration reported the observation of a
relatively narrow enhancement in the $\pi^+\psi'$ invariant mass
distribution in the $B\to K \pi^+\psi'$ decay \cite{pakhlova,4430belle}. This new resonance,
$Z^+(4430)$, is unique among other exotic meson candidates, since it is
the first state which
has a non-zero electric charge. Different theoretical interpretations
were suggested \cite{pakhlova}. Maiani et al. \cite{mprZ} give  qualitative arguments that
the $Z^+(4430)$ could be the first radial excitation ($2S$) of a diquark-antidiquark
$X^+_{u\bar d}(1^{+-}; 1S)$ state ($A\bar A$) with mass 3882 MeV. Our calculations
indicate that the $Z^+(4430)$ can indeed be the $1^{+}$ $2S$
$[cu][\bar c \bar d]$ tetraquark state. It could be the first radial
excitation of the  ground state $(S\bar A- \bar S A)/\sqrt2$, which has the
same mass as $X(3872)$. The other possible interpretation is the $0^{+}$ $2S$
$[cu][\bar c \bar d]$ tetraquark state ($A\bar A$) which has a very
close mass. Measurement of the $Z^+(4430)$ spin will discriminate
between these possibilities. 

Encouraged by this discovery, the Belle Collaboration performed a study
of $\bar B^0\to K^-\pi^+\chi_{c1}$ and observed a
double peaked structure in the $\pi^+\chi_{c1}$ invariant mass
distribution \cite{bellez1z2}. These two charged hidden charm peaks,
$Z(4051)$ and $Z(4248)$, are explicitly exotic. We find no tetraquark
candidates for the former, $Z(4051)$, structure. On the other hand, we
see from Table~\ref{tab:cemass}  that
$Z(4248)$ can be interpreted in our model as the charged partner of
the $1^{-}$ $1P$ state $S\bar S$ or as  the $0^-$  $1P$ state of
the $(S\bar A\pm \bar S A)/\sqrt2$ tetraquark.

In summary, we calculated the masses of excited heavy tetraquarks with hidden
charm in the diquark-antidiquark picture. In contrast to
previous phenomenological treatments, we used the dynamical approach
based on the relativistic quark model. Both diquark and tetraquark
masses were obtained by numerical solution of the quasipotential wave
equation with the corresponding relativistic 
potentials. The diquark structure was taken into account in terms of
diquark wave functions. It is important to emphasize  
that, in our analysis, we did not introduce any free adjustable
parameters but used their values fixed from our previous considerations
of heavy and light hadron properties. It was found that the $X(3872)$, $Z(4248)$,
$Y(4260)$, $Y(4360)$, $Z(4430)$ and $Y(4660)$ exotic meson candidates
can be tetraquark states with hidden charm. 

The authors are grateful to A. Dorokhov, S. Eidelman, I. Ginzburg, V. Matveev,
G. Pakhlova, V. Savrin and O. Teryaev  for support and discussions.  One of us
(D.E.) is grateful to V.Voronov and the other colleagues of BLTP
JINR Dubna for kind hospitality and 
to Bundesministerium f\"ur Bildung and Forschung for financial
support. This work was supported in
part by the Russian Science Support Foundation
(V.O.G.) and the Russian Foundation for Basic Research (RFBR) (grant
No.08-02-00582) (R.N.F. and V.O.G.).

\end{document}